\documentclass[conference]{IEEEtran}
\usepackage{graphicx}
\usepackage{amsfonts}
\usepackage{mathrsfs}
\usepackage{amssymb,amsmath}
\usepackage{bm}
\usepackage{geometry}
\usepackage{geometry}
\usepackage{algorithmic}
\usepackage{theorem}
\usepackage{tikz}
\usepackage[ruled]{algorithm2e}
\usetikzlibrary{shapes.geometric, arrows}
\usepackage{array,color}
\usepackage{cite}
\usepackage{stfloats}
\usepackage{subfigure}
\usepackage{flushend}
\usepackage{array}
\usepackage{multirow}
\usepackage{bbding}
\usepackage{epstopdf}
\newcommand{\tb}{\mathbf}
\newcommand{\rmsg}{\overrightarrow}
\newcommand{\lmsg}{\overleftarrow}

\newlength{\messagetablewidth}
\newcounter{saveequationcntr}%

{\begin{table}%
		\setcounter{saveequationcntr}{\value{equation}}%
		\setcounter{equation}{0}%
	}%
	{\setcounter{equation}{\value{saveequationcntr}}%
\end{table}}
\IEEEoverridecommandlockouts

\begin{document}
	\title{Joint Data and Active User Detection for Grant-free FTN-NOMA in Dynamic Networks}
	\author{\IEEEauthorblockN{Weijie Yuan$^{1}$, Nan Wu$^{2}$, Jinhong Yuan$^{1}$, Derrick Wing Kwan Ng$^{1}$, and Lajos Hanzo$^{3}$}
\IEEEauthorblockA{
$^{1}$School of Electrical Engineering and Telecommunications, University of New South Wales, NSW 2052, Australia\\
$^{2}$School of Information and Electronics, Beijing Institute of Technology, Beijing 100081, China\\
$^{3}$School of Electronics and Computer Science, University of Southampton, SO17 1BJ, UK\\
Email: \{weijie.yuan, j.yuan, w.k.ng\}@unsw.edu.au, wunan@bit.edu.cn, lh@soton.ac.uk}
\thanks{This work is supported the Australia Research Council Discovery Project (DP190101363) and Linkage Projects (LP 160100708 and LP170101196). D. W. K. Ng is supported by funding from the UNSW Digital Grid Futures Institute, UNSW, Sydney, under a cross-disciplinary fund scheme and by the Australian Research Council's Discovery Project (DP190101363). L. Hanzo would like to acknowledge the financial support of the Engineering and Physical Sciences Research Council projects EP/Noo4558/1, EP/PO34284/1, COALESCE, of the Royal Society's Global Challenges Research Fund Grant as well as of the European Research Council's Advanced Fellow Grant QuantCom.}
}	
	
	\maketitle

	\begin{abstract}
Both faster than Nyquist (FTN) signaling and non-orthogonal multiple access (NOMA) are promising next generation wireless communications techniques as a benefit of their capability of improving the system's spectral efficiency. This paper considers an uplink system that combines the advantages of FTN and NOMA. Consequently, an improved spectral efficiency is achieved by deliberately introducing both inter-symbol interference (ISI) and inter-user interference (IUI). More specifically, we propose a grant-free transmission scheme to reduce the signaling overhead and transmission latency of the considered NOMA system. To distinguish the active and inactive users, we develop a novel message passing receiver that jointly estimates the channel state, detects the user activity, and performs decoding. We conclude by quantifying the significant spectral efficiency gain achieved by our amalgamated FTN-NOMA scheme compared to the orthogonal transmission system, which is up to 87.5\%.
		
	\end{abstract}
	
	\section{Introduction}

Wireless communications has played an increasingly important role in modern digital economics. The rapid development of communication technologies has fueled the roll-out of the Internet-of-things (IoT) in 5G wireless systems, which requires to accommodate a massive number of devices \cite{wong2017key}. Unfortunately, current techniques can only support a limited number of active devices concurrently\cite{zeng2019two}. Thus, new techniques supporting massive connectivity are sought.

Recent investigations on non-orthogonal multiple access (NOMA) show that by introducing controllable interference, multiple users can share the same orthogonal radio resources, which allows a communication system to support more users relying on the same amounts of resource elements as OMA \cite{wei2019performanceGainJournal,mohammadkarimi2018signature}. To address the demand for further increasing the spectral efficiency, Faster-than-Nyquist (FTN) signaling, proposed by Mazo in 1970s \cite{mazo1975faster}, has attracted substantial interests, since it can transmit at a symbol-rate beyond the Nyquist rate.

Naturally, gleaning further gains is expected by additionally multiplexing the FTN users employing the popular NOMA principles. As a result, the co-existence of inter-user interference (IUI) as well as inter-symbol interference (ISI) in the FTN-NOMA systems leads to a receiver complexity that grows exponentially with both the number of ISI taps and that of the users. To address this issue, several authors have developed reduced-complexity receivers for FTN signaling \cite{Detection-li2017reduced,sugiura2014frequency} and NOMA \cite{liang2017non,yuan2018iterative}. As a further challenge, the large number of users to be supported by next-generation systems makes the conventional grant-based access control impractical owing to the associated excessive communication overhead and signaling latency\cite{SunZhuo}. To tackle this problem, grant-free schemes dispensing with ``handshaking'' have received considerable attention in NOMA scenarios. For instance, the authors of \cite{zhang2018block} and \cite{yuan2019iterative} assumed the user activity to be static and designed factor graph based receivers for simultaneously solving the channel estimation as well as data and user activity detection problem of NOMA systems. Nevertheless, the user activity in networks fluctuates over time in practice. Some active users may become inactive in the next few time slots, while several sleeping users may become active. Therefore, the identification of user activity in real time is desired.
	
In this paper, we intrinsically amalgamate FTN signaling with the grant-free NOMA concept and consider a dynamic environment, where both the channel and the user activity are time varying. By employing the autoregressive (AR) model of \cite{yuan2020iterative} for approximating the correlated noise samples imposed by FTN signaling, we construct a factor graph and propose a bespoke expectation maximization (EM) - message passing algorithm (MPA) for iteratively estimating both the user activity as well as the channel coefficients, and for detecting the data symbols. Since all messages defined over the factor graph and the solutions of the M-step of the EM are obtained in closed forms, the proposed receiver only has a linearly increasing complexity with respect to the number of users. Our simulation results show that the FTN-NOMA system relying on the proposed receiver significantly improves the spectral efficiency and reliably distinguishes the active/inactive users.

	\emph{Notations:} The superscript $(\cdot)^T$ {and} $(\cdot)^{*}$ denote the transpose and inverse operations, respectively; $\mathcal{G}(\tb{m}_\tb{x},\tb{V}_\tb{x})$ denotes the Gaussian distribution of variable $\tb{x}$ having a mean vector of $\tb{m}_\tb{x}$ and covariance matrix of $\tb{V}_\tb{x}$; $J_0(\cdot)$ denotes the zeroth-order Bessel function of the first kind; $\textrm{diag}\{\tb{a}\}$ denotes a diagonal matrix with the diagonal elements $\tb{a}$; $\mathbb{E}[\cdot]$ denotes the expectation operator; $\propto$ represents both sides of the equation are multiplicatively connected to a constant; $\partial$ denotes the partial derivative operator.

\section{System Model}
The NOMA uplink is considered, where $K$ simultaneous users transmit their information to the BS relying on $J$ orthogonal resource elements, with $J<K$ and $\rho=\frac{K}{J}$ representing the normalized user-load. The coded bit stream corresponding to the $k$th user is first mapped to a sequence of data symbols and then spread over $J$ resource-slots using a low-density signature (LDS) $\tb{x}_k^{[n]}=[x_{k,1}^{[n]},...,x_{k,J}^{[n]}]^T$, where $x_{k,j}^{[n]}$ denotes the symbol of user $k$ occupying the $j$th resource element at time instant $n$.
To employ FTN signaling in the NOMA system, the transmitted sequences, $\tb{x}_k^{[n]}$ of different users pass through a shaping filter $q(t)$, having a symbol period\footnote{We assume that for all users, the same shaping filter $q(t)$ and packing factor $\tau$ are employed.} of $\tau T_0$, yielding
\begin{align}
	\tb{s}_k(t)=\sum_n \tb{x}_k^{[n]} q(t-n\tau T_0),
\vspace{0mm}
\end{align}
where $\tau$ is the FTN packing factor \cite{Detection-li2017reduced} and $\tb{s}_k(t)=[s_{k,1}(t),...,s_{k,J}(t)]^T$. The transmitted signals of all users are multiplexed over $J$ resource elements and passed through a time-variant fading channel $\tb{h}_k(t)=[h_{k,1}(t),...,h_{k,J}(t)]^T$. Assuming perfect synchronization between the BS and the users, the signal received at the BS obeys:
\begin{align}
	\tb{y}(t)=\sum_{k=1}^K \textrm{diag}\{\tb{h}_k(t)\} \tb{s}_k(t) +\tb{w}(t),
\end{align}
where $\tb{y}(t)$ and $\tb{w}(t)$ are both $J$-dimensional vectors with the $j$th entries being the received signal and noise at the $j$th resource element, respectively. We now introduce a binary variable $\lambda_k^{[n]}=\{0,1\}$ denoting the user-activity, where $\lambda_k^{[n]}=1$ represents an active user and 0 an inactive one. Then after processing by a matched filter $q^{*}(-t)$, the discrete time model for the received signal is given by
\begin{align}\label{rjn_random_access}
	{r}_{j}^{[n]}=\sum_{k=1}^K \lambda_k^{[n]} h_{k,j}^{[n]} \sum _{i=-L}^L g_i x^{[n-i]}_{k,j}+\omega_{j}^{[n]},
\end{align}
where $g_i$ denotes the FTN signaling-induced ISI tap and $L$ is the length of the taps.
Note that in FTN signaling, the noise samples of different time slots are correlated, which imposes challenge on the receiver design. As a remedy, we employ an AR model to approximate the colored noise samples. In practice, the AR model with an AR order of $n_b$ is used for approximating the noise sample $\omega_j^{[n]}$, given by
\begin{align}\label{omega_arma}
\vspace{0mm}
\omega_j^{[n]}\approx e_j^{[n]} +\sum_{i=1}^{n_b} b_i  \omega_j^{[n-i]},
\vspace{0mm}
\end{align}
where $e_j^{[n]}$ is a random Gaussian impairment having zero-mean as well as variance $\sigma_e^2$ and $b_i$ is the AR coefficient. Based on the known coefficients $\{g_i\}$, the parameters $\{b_i\}$ and $\sigma_e^2$ can be estimated by solving the Yule-Walker equations \cite{levy2008principles}.

%

 \section{The Proposed Low-complexity Receiver Design}
 From a statistical inference perspective, inferring channel coefficients, and all users' information bits from the received signal samples is equivalent to determining the \emph{a posteriori} distributions of the corresponding variables.
 \subsection{Factor Graph Representation}
By stacking all transmitted symbols, received samples, channel taps, user states and noise samples into vectors, i.e., $\tb{x}$, $\tb{r}$, $\tb{h}$, $\bm{\lambda}$, and $\bm{\omega}$, the joint \emph{a posteriori} distribution is written as $p(\tb{x},\tb{h},\bm{\lambda},\bm{\omega}|\tb{r})$. For an unknown variable $z$, we aim for deriving its marginal distribution $p(z|\tb{r})$ and estimating it via the maximum \emph{a posteriori} (MAP) estimators formulated as
\begin{align}
	z = \arg\max_z p(z|\tb{r}).
\end{align}
Nevertheless, direct marginalization is usually intractable due to the associated high-dimensional integration. As an alternative, the factor graph framework is capable of circumventing this problem by exploiting the conditional independence of variables given the observations. Exploiting the fact that $\tb{x}$, $\tb{h}$, $\bm{\lambda}$, and $\bm{\omega}$ are independent of each other, $p(\tb{x},\tb{h},\bm{\lambda},\bm{\omega}|\tb{r})$ can be factorized as
\begin{align*}
	p(\tb{x},\tb{h},\bm{\lambda},\bm{\omega}|\tb{r})\propto p(\tb{x})\cdot p(\tb{h}) \cdot p(\bm{\lambda})\cdot p(\bm{\omega}) \cdot p(\tb{r}|\tb{x},\tb{h},\bm{\lambda},\bm{\omega}).
\end{align*}
Since the transmitted symbols of different users at different instants are independent, $p(\tb{x})$ can be fully factorized as
\begin{align}\label{factor_x}
	p(\tb{x})=\prod_{k,j,n} p(x_{k,j}^{[n]}),
\end{align}
where $p(x_{k,j}^{[n]})$ is determined based on the log-likelihood ratios (LLRs) output by the channel decoder.

For time-varying channel taps $\tb{h}$, it is convenient to characterize $h_{k,j}^{[n]}$ by a Gauss-Markov model $h_{k,j}^{[n]}=\alpha h_{k,j}^{[n-1]}+\varepsilon^{[n]}$, where the coefficient $\alpha$ obeys the zero-order Bessel function of the first kind\cite{hanzo2011mimo}
\begin{align}
	\alpha=\mathbb{E}[h_{k,j}^{[n]} (h_{k,j}^{[n]})^{*}]=J_0 (2\pi f_D \tau T_0),
\end{align}
and $\varepsilon^{[n]}$ is a zero-mean Gaussian distributed variable with variance $1-|\alpha|^2$. Consequently, $p(\tb{h})$ is expressed as
\begin{align}
	p(\tb{h})=\prod_{k,j} p(h_{k,j}^0)\prod_n {p(h_{k,j}^{[n]}|h_{k,j}^{[n-1]})}.
\end{align}

In a dynamically fluctuating environment, the evolution of the user-activity state $\lambda$ can be modeled by a Markov chain, where the current activities of the users depend on the states of the previous time instant. Hence, the distribution $p(\bm{\lambda})$ can be factorized as
\begin{align}
	p(\bm{\lambda})=\prod_{k=1}^K p(\lambda_k^{0})\cdot \prod_n p(\lambda_k^{[n]}|\lambda_k^{[n-1]}).
\end{align}
Depending on the previous state of user $k$, the state transition function $p(\lambda_k^{[n]}|\lambda_k^{[n-1]})$ has different expressions. Given the user-birth probability of $p_{b_k}^{[n]}$ and the mortality probability of $p_{m_k}^{[n]}$, the transition probability of user-activity state $p(\lambda_k^{[n]}|\lambda_k^{[n-1]})$ is expressed as
\begin{align}
	p(\lambda_k^{[n]}|\lambda_k^{[n-1]}) = \left\{\begin{array}{cc}
		(p_{b_k}^{[n]})^{1-\lambda_k^{[n-1]}} (1-p_{m_k}^{[n]})^{\lambda_k^{[n-1]}} &\lambda_k^{[n]}=1,\\
		(1-p_{b_k}^{[n]})^{1-\lambda_k^{[n-1]}} (p_{m_k}^{[n]})^{\lambda_k^{[n-1]}} &\lambda_k^{[n]}=0.
	\end{array}
	\right.
\end{align}
Assuming that $\Lambda$ denotes the average number of users becoming active at a time instant, we set $p_{b_k}^{[n]}=\Lambda/K$ as the birth probability of a user. For the mortality probability, establishing an accurate model requires a large amount of data, but this is beyond the scope of the paper. Hence we employ a fair scheme assuming that $p_{m_k}^{[n]}=0.5$.

Based on the AR model \eqref{omega_arma} of the noise sample $\omega_j^{[n]}$, we can factorize $p(\bm{\omega})$ as
\begin{align}
	p(\bm{\omega})=\prod_j p(\omega_j^0)\prod_n {p(\omega_j^{[n]}|\omega_j^{[n-n_b]},...,\omega_j^{[n-1]})},
\end{align}
where $p(\omega_j^0)\propto \mathcal{G}(0,\sigma_{e}^2)$ and $p(\omega_j^{[n]}|\omega_j^{[n-n_b]},...,\omega_j^{[n-1]})\propto \exp\left(-\frac{(\omega_j^{[n]}-\tb{b}^T \bm{\omega}_j^{[n]})^2}{\sigma_{e}^2}\right)$, with the notations $\tb{b}=[b_1,...,b_{n_b}]^T$ and $\bm{\omega}_j^{[n]}=[\omega_{j}^{[n-1]},...,\omega_{j}^{[n-n_b]}]^T$. Furthermore, we can write the evolution model of $\bm{\omega}_j^{[n+1]}$ as
\begin{align}\label{omega_evolve}
	\bm{\omega}_j^{[n+1]} = \tb{B}_1\bm{\omega}_j^{[n]} + \tb{b}_1 \omega_j^{[n]},
\end{align}
where $\tb{B}_1=\left[\begin{array}{cc}
	\tb{0}_{n_b-1}^T&0\\
	\tb{I}_{n_b-1}&\tb{0}_{n_b-1}
\end{array}
\right]$ and $\tb{b}_1=[1,\tb{0}_{n_b-1}^T]^T$.

Based on \eqref{rjn_random_access}, we use a Dirac delta function $\delta(\cdot)$ for representing the relationship between the received signal sample and the unknown variables. By introducing an auxiliary variable\footnote{The auxiliary variable is introduced to reduce the number of multiplications \cite{yuan2018iterative}.} $s_{k,j}^{[n]}=\sum_{i=-L}^L g_i x_{k,j}^{[n-i]}=\tb{g}^T \tilde{\tb{x}}_{k,j}^{[n]}$, $p(\tb{r}|\tb{x},\tb{h},\bm{\lambda},\bm{\omega})$ is factorized as
\begin{align}\label{factor_likelhoood}
	p(\tb{r}|\tb{x},\tb{h},\bm{\lambda},\bm{\omega}) = \prod_{j,n}& {\delta (r_j^{[n]}-\sum_{k=1}^K \lambda_k^{[n]} h_{k,j}^{[n]} s_{k,j}^{[n]}-\omega_{j}^{[n]})}\nonumber\\&\cdot {\delta(s_{k,j}^{[n]}-\tb{g}^T \tilde{\tb{x}}_{k,j}^{[n]})},
\end{align}
 where $\tb{g}$ and $\tilde{\tb{x}}_{k,j}^{[n]}$ denote the vectors $[g_{-L},...,g_L]^T$ and $[{{x}}_{k,j}^{[n+L]},...,{{x}}_{k,j}^{[n-L]}]$, respectively.
The variable $\tilde{\tb{x}}_{k,j}^{[n]}$ follows a similar evolution model as in \eqref{omega_evolve},
\begin{align}
	\tilde{\tb{x}}_{k,j}^{[n]}=\tb{B}_2 \tilde{\tb{x}}_{k,j}^{[n-1]}+\tb{b}_2 x_{k,j}^{[n+L]},
\end{align}
where $\tb{B}_2=\left[\begin{array}{cc}
	\tb{0}_{{2L}}^T&0\\
	\tb{I}_{{2L}}&\tb{0}_{{2L}}
\end{array}
\right]$ and $\tb{b}_2=[1,\tb{0}_{{2L}}^T]^T$.

\begin{figure}
	\centering
	\subfigure[Subgraph for the multiuser detection and decoding part.]{\label{fg_part}
\includegraphics[width=.45\textwidth]{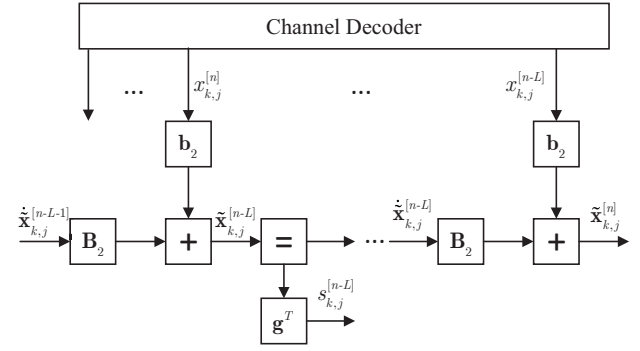}}
	\subfigure[Factor graph representation of the FTN-NOMA system. The notations $\phi_k^{[n]}$, $\psi_{k,j}^{[n]}$, and $\Omega_j^{[n]}$ represent the function $p(\lambda_k^{[n]}|\lambda_k^{[n-1]})$, $p(h_{k,j}^{[n]}|h_{k,j}^{[n-1]})$, and $p(\omega_j^{[n]}|\omega_j^{[n-n_b]},...,\omega_j^{[n-1]})$.]{\label{fg_whole}
\includegraphics[width=.5\textwidth]{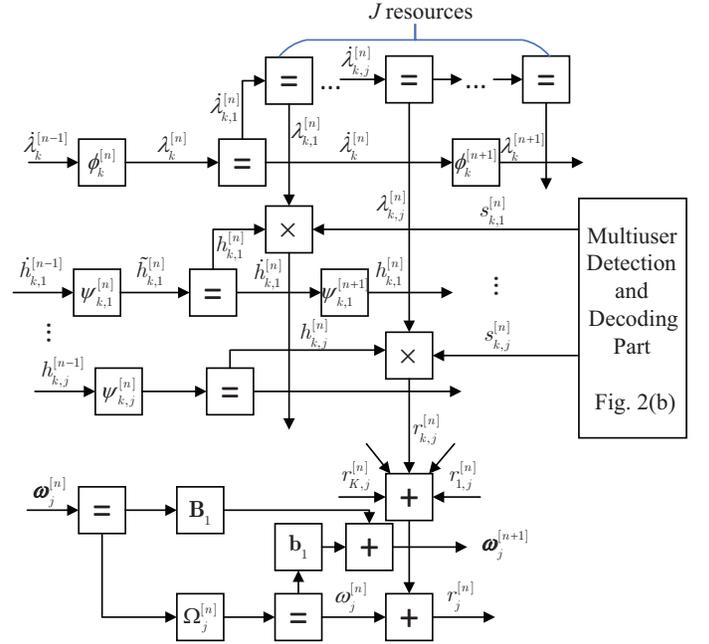}}
	\caption{Factor graph for joint user activity tracking and data detection. }
	\label{FG_random_access}
	\centering
\end{figure}

Based on the factorizations of \eqref{factor_x}-\eqref{factor_likelhoood}, we now have the factorization of $p(\tb{x},\tb{h},\bm{\lambda},\bm{\omega}|\tb{r})$, which we represent by a factor graph, as depicted in Fig. \ref{FG_random_access}. On this factor graph, the factor nodes denoted by squares represent the functions nodes while the variables are denoted by edges. The equality factor nodes of Fig. 1 represented by the symbol $\framebox{$=$}$ are introduced for variable ``cloning'' to enforce the condition that a variable may only appear in a maximum of two functions. To simplify the notations, we adopt $\rmsg{\mu}(x)$ and $\lmsg{\mu}(x)$ to denote the specific messages of the variable $x$ that flow in the direction and in the opposite direction of the edge.

\subsection{MPA for Message Calculations}
\subsubsection{Multiuser Detection and Decoding Part}
Let us commence from the message calculations in Fig. 1(a). The intrinsic information $\rmsg{\mu}(x_{k,j}^{[n]})$ is calculated based on the log-likelihood ratio (LLR) output by the channel decoder. Since $x_{k,j}^{[n]}$ is discretely distributed, the message passing algorithm (MPA) exhibits an excessive complexity. To this end, we approximate $\rmsg{\mu}(x_{k,j}^{[n]})$ by a Gaussian distribution using expectation propagation (EP). Assuming that the extrinsic message $\lmsg{\mu}(x_{k,j}^{[n]})$ obeys the Gaussian distribution of $\mathcal{G}(\lmsg{m}_{x_{k,j}^{[n]}},\lmsg{v}_{x_{k,j}^{[n]}})$, we can readily obtain the mean and variance of $b(x_{k,j}^{[n]})$ by moment matching and then determine the Gaussian belief $\tilde{b}(x_{k,j}^{[n]})\propto \mathcal{G}({m}_{x_{k,j}^{[n]}},{v}_{x_{k,j}^{[n]}})$. Hence we have the message $\rmsg{\mu}(x_{k,j}^{[n]})$ expressed as \begin{align}
\rmsg{\mu}(x_{k,j}^{[n]})={\tilde{b}(x_{k,j}^{[n]})}/{\lmsg{\mu}(x_{k,j}^{[n]})}\propto\mathcal{G}(\rmsg{m}_{x_{k,j}^{[n]}},\rmsg{v}_{x_{k,j}^{[n]}}).
\end{align}

By employing the MPA rules, we can determine the mean vector and covariance matrix of the message $\rmsg{\mu}(\dot{\tb{x}}_{k,j}^{[n-L]})$. Therefore the message output by the multiuser detector can be assumed to be Gaussian, formulated as
\begin{align}
\rmsg{\mu}(s_{k,j}^{[n-L]})\propto \mathcal{G}(\tb{g}_T\rmsg{\tb{m}}_{\dot{\tb{x}}_{k,j}^{[n-L]}},\tb{g}_T\rmsg{\tb{V}}_{\dot{\tb{x}}_{k,j}^{[n-L]}}\tb{g}).
\end{align}
Finally, we are interested in calculating the extrinsic message $\lmsg{\mu}(x_{k,j}^{[n]})$, whose mean and variance obey
\begin{align}
\lmsg{m}_{x_{k,j}^{[n]}}&=\tb{b}_2^T\left(\lmsg{\tb{m}}_{\tilde{\tb{x}}_{k,j}^{[n-L]}}-\tb{B}_2 \rmsg{\tb{m}}_{\dot{\tilde{\tb{x}}}_{k,j}^{[n-L-1]}}\right),\\
\lmsg{v}_{x_{k,j}^{[n]}}&=\tb{b}_2^T\left(\lmsg{\tb{V}}_{\tilde{\tb{x}}_{k,j}^{[n-L]}}+\tb{B}_2 \rmsg{\tb{V}}_{\dot{\tilde{\tb{x}}}_{k,j}^{[n-L-1]}}\tb{B}_2 ^T\right)\tb{b}_2,
\end{align}
respectively.

\subsubsection{Channel Estimation}
Provided that the message $\rmsg{\mu}(\dot{h}_{k,j}^{[n-1]})$ is available in Gaussian form, by applying the MPA rules, the message $\rmsg{\mu}(\tilde{h}_{k,j}^{[n]})$ is expressed as
\begin{align}\label{hpredict}
 \rmsg{\mu}(\tilde{h}_{k,j}^{[n]})&\propto \mathcal{G}\left(\rmsg{m}_{\tilde{h}_{k,j}^{[n]}},\rmsg{v}_{\tilde{h}_{k,j}^{[n]}}\right)\nonumber\\
&\propto\left(\alpha \rmsg{m}_{\dot{h}_{k,j}^{[n-1]}},1+|\alpha|^2(\rmsg{v}_{\dot{h}_{k,j}^{[n-1]}}-1)\right).
\end{align}
Then, we are able to derive the message $\rmsg{\mu}(\dot{h}_{k,j}^{[n]})$, given by 
\begin{align}\label{weightmat}
  \rmsg{v}_{\dot{h}_{k,1}^{[n]}}&=\frac{\rmsg{v}_{\tilde{h}_{k,1}^{[n]}}+\lmsg{v}_{{h}_{k,1}^{[n]}}}{\rmsg{v}_{\tilde{h}_{k,1}^{[n]}}\lmsg{v}_{{h}_{k,1}^{[n]}}}, \\\label{transformedmean}
  \rmsg{m}_{\dot{h}_{k,1}^{[n]}}&=\rmsg{v}_{\dot{h}_{k,1}^{[n]}}\left(\frac{\rmsg{m}_{\tilde{h}_{k,1}^{[n]}}}{\rmsg{v}_{\tilde{h}_{k,1}^{[n]}}}+\frac{\lmsg{m}_{{h}_{k,1}^{[n]}}}{\lmsg{v}_{{h}_{k,1}^{[n]}}}\right).
\end{align}

Next, we consider the process of colored noise. Since the AR process given by \eqref{omega_arma} is causal, the messages are only propagated forward along the arrow's direction. Provided that the means of the noise parameters are $0$, we can readily derive the corresponding messages as follows
\begin{align}\label{colorednoise1}
  \rmsg{v}_{\omega_j^{[n]}}  &= \tb{b}^T\rmsg{\tb{V}}_{\bm{\omega}_j^{[n]}}\tb{b}, \\
  \rmsg{\tb{V}}_{\bm{\omega}_j^{[n+1]}} & = \tb{B}_1  \rmsg{\tb{V}}_{\bm{\omega}_j^{[n]}} \tb{B}_1^T +\tb{b}_1 \rmsg{v}_{\omega_j^{[n]}}\tb{b}_1^T.
\end{align}


\subsubsection{User-activity Idenfitication}
For the discrete random variable representing a user-activity state at time instant $n-1$, we have $\dot{\lambda}_{k}^{[n-1]}$, and the message $\rmsg{\mu}(\dot{\lambda}_{k}^{[n-1]})$ is the belief of user $k$'s state at time instant $n-1$, which is fully characterized by the probability of $\lambda_k^{[n-1]}=1$, i.e., $\rmsg{p}_{\dot{\lambda}_k^{[n-1]}}$.
Therefore, passing on the user-activity probability $\rmsg{p}_{\dot{\lambda}_k^{[n-1]}}$ instead of the message can simplify the expressions. We arrive at the forward message $\rmsg{p}(\lambda_k^{[n]})$ expressed as
\begin{align}
\rmsg{p}_{\lambda_k^{[n]}}=(1-p_{m_k}^{[n]})\rmsg{p}_{\dot{\lambda}_k^{[n]}}+p_{b_k}^{[n]}(1-\rmsg{p}_{\dot{\lambda}_k^{[n]}}).
\end{align}
The equality node of Fig. 1 is equivalent to the product of messages. Therefore the message updating concerning $\dot{\lambda}_{k}^{[n]}$ is derived as
\begin{align}
  \rmsg{p}_{\dot{\lambda}_{k}^{[n]}}=\frac{\rmsg{p}{\lambda_k^{[n]}} \lmsg{p}_{\dot{\lambda}_{k,1}^{[n]}}}{1-\rmsg{p}{\lambda_k^{[n]}}-\lmsg{p}_{\dot{\lambda}_{k,1}^{[n]}}}.
\end{align}
The message $\rmsg{p}_{{\lambda}_{k,j}^{[n]}}$ forwarded to the multiplier node can be obtained similarly.

Note that the above message calculations depend on the assumption of having known backward messages gleaned from the multiplier node. According to the update rules of the conventional MPA, the messages derived at the multiplier node \framebox{$\times$} are unable to provide Gaussian form messages. Hence, we invoke the expectation maximization algorithm for the multiplier node.

\subsection{Modified EM Algorithm for {$\times$} Node}\label{EM111}
Without loss of generality, we consider the multiplier node connected with $r_{k,j}^{[n]}$ and the joint distribution $p({\lambda}_{k,j}^{[n]},{s}_{k,j}^{[n]},{h}_{k,j}^{[n]}|{r}_{k,j}^{[n]})$. We first define $\lambda_{k,j}^{[n]}$ as the unknown parameter and $\{h_{k,j}^{[n]}, s_{k,j}^{[n]},r_{k,j}^{[n]}\}$ as the complete data set associated with incomplete data $r_{k,j}^{[n]}$ and latent variables $h_{k,j}^{[n]}$, $s_{k,j}^{[n]}$. 
Assuming that the beliefs $b(s_{k,j}^{[n]})$ and $b(h_{k,j}^{[n]})$ are available, the expectation of the complete-data log augmented density is calculated as
\begin{align}\label{qfunctionlambda}
q({\lambda}_{k,j}^{[n]})
= -\int\int &\frac{\left(\lmsg{m}_{r_{k,j}^{[n]}}-\lambda_{k,j}^{[n]}{s}_{k,j}^{[n]}{h}_{k,j}^{[n]}\right)^2}{\lmsg{v}_{r_{k,j}^{[n]}}}\cdot b(s_{k,j}^{[n]}) b(h_{k,j}^{[n]})\nonumber\\&\textrm{d} h_{s,j}^{[n]}\textrm{d} h_{k,j}^{[n]}+\ln\rmsg{\mu}({\lambda}_{k,j}^{[n]}) +C,
  \end{align}
where $C$ is a constant that is irrelevant to $\lambda_{k,j}^n$. It can be observed that \eqref{qfunctionlambda} is a concave function and the estimate $\hat{\lambda}_{k,j}^{[n]}$ is given by the solution of $\frac{\partial q({\lambda}_{k,j}^{[n]})}{\partial {\lambda}_{k,j}^{[n]}}=0$. However, note that \eqref{qfunctionlambda} only considers the $j$th resource element, while the user activity applies to all radio resources. The maximization should be performed by obtaining the necessary information from all $J$ resource elements, i.e., with respect to the variable $\dot{\lambda}_{k}^n$. To this end, the multiplier node will feed back the message $\lmsg{\mu}({\lambda}_{k,j}^{[n]})$ to the equality node of Fig. \ref{fg_whole}. Having $q({\lambda}_{k,j}^{[n]})$ in hand, $\lmsg{\mu}({\lambda}_{k,j}^{[n]})$ is calculated as $\exp\left[q({\lambda}_{k,j}^{[n]})\right]/{\rmsg{\mu}({\lambda}_{k,j}^{[n]})}$.
Similar to the forward message $\rmsg{p}(\lambda_k^{[n]})$, we can simply pass on the normalized probability $\lmsg{p}_{{\lambda}_{k,j}^{[n]}}$,
which is used to calculate $\lmsg{p}_{{\lambda}_{k,j-1}^{[n]}}$, then $\lmsg{p}_{{\lambda}_{k,1}^{[n]}}$ and finally $\rmsg{p}_{\dot{\lambda}_k^{[n]}}$.

To obtain the beliefs of $s_{k,j}^{[n]}$ and $h_{k,j}^{[n]}$ , we apply the concept of EM again that $s_{k,j}^{[n]}$ becomes an unknown parameter and $h_{k,j}^{[n]}$ remains the latent variable. In this way, the belief of $s_{k,j}^{[n]}$ is updated as follows:
\begin{align}\label{bskjn}
b(s_{k,j}^{[n]})\propto\rmsg{\mu}(s_{k,j}^{[n]})\cdot \exp\Big(\int &b(h_{k,j}^{[n]})\ln p(r_{k,j}^{[n]}, s_{k,j}^{[n]}|h_{k,j}^{[n]},\hat{\lambda}_{k,j}^{[n]}) \nonumber\\&\textrm{d} h_{k,j}^{[n]}\Big),
\end{align}
where $\lambda_{k,j}^{[n]}$ is replaced by the estimate $\hat{\lambda}_{k,j}^{[n]}$ obtained from the maximization of \eqref{qfunctionlambda}, expressed as
\begin{align}
  \hat{\lambda}_{k,j}^{[n]}=\frac{\lmsg{m}_{r_{k,j}^{[n]}}m_{s_{k,j}^{[n]}}m_{h_{k,j}^{[n]}}
 +\lmsg{v}_{r_{k,j}^{[n]}}(1-2\rmsg{p}_{\lambda_{k,j}^{[n]}})}{(|m_{s_{k,j}^{[n]}}|^2+v_{s_{k,j}^{[n]}})(|m_{h_{k,j}^{[n]}}|^2+v_{h_{k,j}^{[n]}})}.
\end{align}
Since we have $b(s_{k,j}^{[n]})=\rmsg{\mu}(s_{k,j}^{[n]})\cdot \lmsg{\mu}(s_{k,j}^{[n]})$, it is natural to define the second term on the right-hand side of \eqref{bskjn} as $\lmsg{\mu}(s_{k,j}^{[n]})$. Assuming that $b(h_{k,j}^{[n]})$, it follows that $\mathcal{G}(m_{h_{k,j}^{[n]}},v_{h_{k,j}^{[n]}})$, $\lmsg{\mu}(s_{k,j}^{[n]})$ can be modeled by a Gaussian distribution with a mean and variance of
\begin{align}
  \lmsg{m}_{s_{k,j}^{[n]}}&= \frac{\lmsg{m}_{r_{k,j}^{[n]}} m_{h_{k,j}^{[n]}}}{|m_{h_{k,j}^{[n]}}|^2+v_{h_{k,j}^{[n]}}},\\
    \lmsg{v}_{s_{k,j}^{[n]}}&= \frac{\lmsg{v}_{r_{k,j}^{[n]}} }{|m_{h_{k,j}^{[n]}}|^2+v_{h_{k,j}^{[n]}}}.
  \end{align}
Consequently the belief $b(s_{k,j}^{[n]})$ is readily obtained. By exchanging the roles of $s_{k,j}^{[n]}$ and $h_{k,j}^{[n]}$, we have the updating rules of the message $\lmsg{\mu}(h_{k,j}^{[n]})$ and of the belief $b(h_{k,j}^{[n]})$.
After obtaining the beliefs $b(s_{k,j}^{[n]})$ and $b(h_{k,j}^{[n]})$, we can now determine $q({\lambda}_{k,j}^{[n]})$ in the next iteration following \eqref{qfunctionlambda}.

Finally, we can obtain the forward message $\rmsg{\mu}(r_{k,j}^{[n]})$. Since the variables $\lambda_{k,j}^{[n]}$, $s_{k,j}^{[n]}$, and $h_{k,j}^{[n]}$ are independent, the moments of $r_{k,j}^{[n]}$ are given by the product of the moments of the above three variables. Consequently, we have
\begin{align}\label{rmrkjn}
  \rmsg{m}_{r_{k,j}^{[n]}}=&\hat{\lambda}_{k,j}^n m_{h_{k,j}^{[n]}} m_{h_{s,j}^{[n]}},\\\label{rvrkjn}
  \rmsg{v}_{r_{k,j}^{[n]}}=&\hat{\lambda}_{k,j}^n\big((1-\hat{\lambda}_{k,j}^n)|m_{h_{k,j}^{[n]}}|^2|m_{s_{k,j}^{[n]}}|^2+|m_{h_{k,j}^{[n]}}|^2 v_{s_{k,j}^{[n]}}\nonumber\\&+|m_{s_{k,j}^{[n]}}v_{h_{k,j}^{[n]}}+v_{h_{k,j}^{[n]}}v_{s_{k,j}^{[n]}}|^2\big).
\end{align}
Above, we have obtained $\rmsg{p}_{\dot{\lambda}_k^{[n]}}$ and $b(h_{k,j}^{[n]})$. Then we can proceed by comparing $\rmsg{p}_{\dot{\lambda}_k^{[n]}}$ to a specific threshold for deciding whether the user $k$ is active, while the estimate of the channel tap $h_{k,j}^{[n]}$ is given by $\hat{h}_{k,j}^{[n]}= \mathbb{E}_{h}\left[b(h_{k,j}^{[n]})\right]=m_{h_{k,j}^{[n]}}$.

It is observed that in the proposed algorithm, the messages are represented only by a few parameters and the integration is simplified to additions and multiplications, which dramatically reduces the receiver's complexity. Explicitly, the complexity only increases linearly with the number of active users $|\mathcal{K}^{+}|$, which shows the superiority of the proposed algorithm in massive connectivity scenarios.

	\section{Simulation Results}

This section presents our simulation results. A rate-1/2 length-$2,160$ low-density parity-check (LDPC) code is adopted. At total of $K=180$ users are supported by $J=120$ resource elements in our NOMA system, leading to the normalized user-load of $\rho=150\%$. Quadrature Phase Shift Keying (QPSK) is used for bit-to-symbol mapping. For each user, 5 sequences of symbols having a length of $1,080$ are transmitted. The data symbols corresponding to different users are shaped by the raised root cosine filter having a roll-off factor of $\alpha=0.4$ and a FTN packing factor of $\tau=0.8$ and. A Rayleigh fading channel is considered and the taps are generated by Jake's model with a normalized Doppler rate of $f_D\tau T_0 = 0.005$. The channel estimation is based on the least square (LS) method and as few as 5 pilots. The parameter $\Lambda$ is set to $20$, which indicates that approximately 11\% of users are active. Again, the mortality probability was set to $p_{m_k}^{[n]}=0.5$. The user activity is assumed to remain static for a sequence of 1,080 symbols. The threshold of $0.5$ is employed for user activity identification.

 \begin{figure}[]
	\centering
	\includegraphics[width=.45\textwidth]{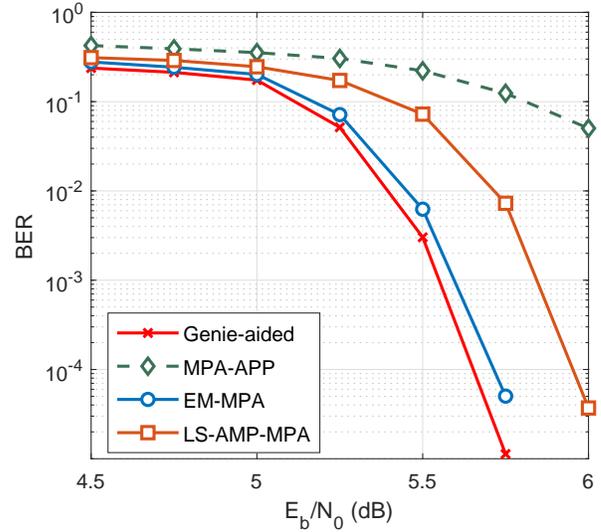}
	\caption{BER performance of the proposed algorithm versus $E_b/N_0$.}\label{fig3}
	\centering
\vspace{-1mm}
\end{figure}
In Fig. \ref{fig3}, we compare the proposed receiver design to some existing benchmark algorithms in terms of its BER performance. Explicitly, the BER performance versus $E_b/N_0$ of the proposed algorithm as well as of the MPA-APP and LS-AMP-MPA methods are illustrated in Fig. \ref{fig3}. The LS-AMP-MPA is a two-step method which firstly identifies the active users and then performs MPA based MUD. Since the two-stage method only provides the estimates of user-activities for data detection, considerable performance loss can be seen compared to the proposed EM-MPA algorithm. The MPA-APP method regards all users to be active although most users are inactive, hence leading to certain performance degradation. Moreover, we also include the curve corresponding to the OMA system using MPA for channel estimation and user-activity identification (Genie-aided). A slight performance loss is observed for the proposed algorithm due to the ISI and IUI imposed by our FTN-NOMA system. Nevertheless, the spectral efficiency is increased by $(\rho-1)*1/\tau=87.5\%$, given the same radio resources.

 \begin{figure}[]
	\centering
	\includegraphics[width=.45\textwidth]{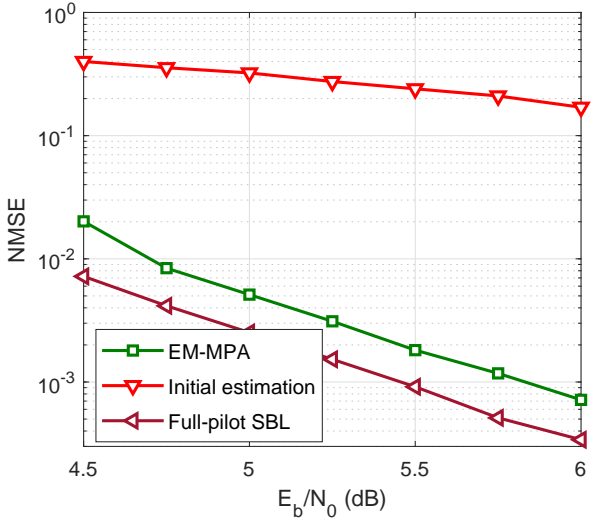}
	\caption{NMSE of channel estimation versus $E_b/N_0$.}\label{fig6}
	\centering
\vspace{-2mm}
\end{figure}
We plot the normalized minimum mean-squared errors (NMSE) of the channel estimate based on the proposed algorithm as well as on the GA-MPA and on the full pilot based sparse Bayesian learning (SBL) method in Fig. \ref{fig6}. Since EP can exploit the extrinsic information from the detector when approximating the discrete distribution by a Gaussian one, the proposed EM-MPA algorithm outperforms the GA-MPA method using direct moment matching. It is also worth noting that the NMSE performance of the proposed algorithm has a modest performance loss compared to the full-pilot based method, which demonstrates the powerful capability of the proposed algorithm.

 \begin{figure}[]
	\centering
	\includegraphics[width=.45\textwidth]{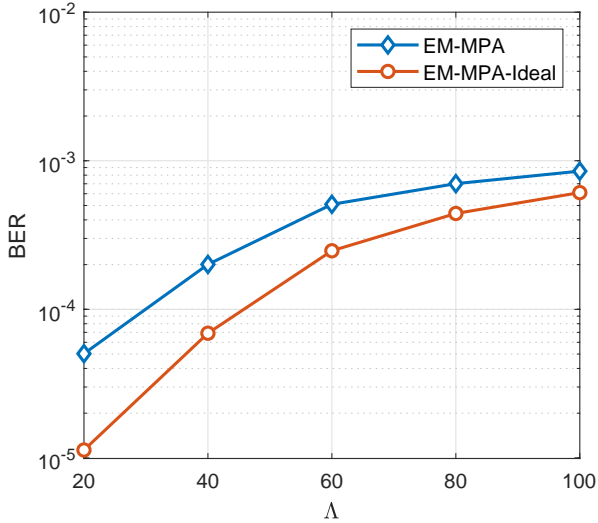}
	\caption{Impact of number of active users on BER performance.}\label{fig8}
	\centering
\vspace{-2mm}
\end{figure}
In Fig. \ref{fig8}, the impact of the number of active users on the decoding performance is considered. We plot the BER performance versus the parameter $\Lambda$ at $E_b/N_0=5.75$ dB, where the performance of the EM-MPA-Ideal relying on perfectly known user-activity is also shown as a performance upper bound. In fact, a higher $\Lambda$ results in more active users. Hence, the IUI becomes more severe and degrades the BER performance, which can be observed for both the proposed method and the performance upper bound. It is interesting to see that when $\Lambda$ increases, the performance of the proposed algorithm approaches the upper bound, because user identification becomes more accurate for more active users. In particular, in the extreme case, when all users are active, the identification error will drop to $0$.


\section{Conclusions}
We have conceived a low-complexity receiver for the grant-free FTN-NOMA uplink. We considered dynamically fluctuating environments, which assume that the user state and channel vary with time. A new expectation maximization - message passing algorithm combination was proposed based on the factor graph framework for joint FTN symbol detection, channel estimation and user-activity tracking. The complexity of the proposed receiver increases linearly with the number of active users, which is significantly lower than that of the conventional message passing receiver. Our simulation results demonstrated the efficiency of the proposed method in identifying active users and decoding the information bits whilst enhancing the bandwidth efficiency by up to 87.5\%.

\bibliographystyle{IEEEtran}
\bibliography{dynamic}
\end{document}